# (1) Overview

## Title

archNEMESIS: an open-source Python package for analysis of planetary atmospheric spectra

## Paper Authors

1. Alday, Juan;
2. Penn, Joseph;
3. Irwin, Patrick;
4. Mason, Jonathon;
5. Yang, Jingxuan;

## Paper Author Roles and Affiliations

1. School of Physical Sciences, The Open University, Milton Keynes, United Kingdom.
2. AOPP, Department of Physics, University of Oxford, Oxford, United Kingdom.
3. AOPP, Department of Physics, University of Oxford, Oxford, United Kingdom.
4. School of Physical Sciences, The Open University, Milton Keynes, United Kingdom.
5. AOPP, Department of Physics, University of Oxford, Oxford, United Kingdom.

## Abstract

ArchNEMESIS is an open-source Python package developed for the analysis of remote sensing spectroscopic observations of planetary atmospheres. It is based on the widely used NEMESIS radiative transfer and retrieval tool, which has been extensively used for the investigation of a wide variety of planetary environments. The main goal of archNEMESIS is to provide the capabilities of its Fortran-based predecessor, keeping or exceeding the efficiency in the calculations, and benefitting from the advantages Python tools provide in terms of usability and portability. The code, stored in a public GitHub repository under a GPL-v3.0 license, is accompanied by detailed documentation available at https://archnemesis.readthedocs.io/.

## 1.1. Introduction

Remote sensing is one of the most powerful techniques to investigate the state and composition of planetary atmospheres, including those of Solar System planets and exoplanets, providing crucial information to our understanding of planet formation, atmospheric physics and the potential for habitability. In the present stage of space exploration, the volume and resolution of remote sensing data is unprecedented, and interpreting such vast datasets requires the development of advanced, efficient analysis tools. Moreover, because of the wide variety of atmospheres that are currently studied, and the many different observation types (e.g., space or ground-based telescope observations, nadir or limb-viewing orbiters, entry probes), a flexible computational framework that can be adapted for different environments and geometries is highly desirable.

The NEMESIS (Non-linear Optimal Estimator for MultivariatE Spectral analySIS) code [1] was developed as a general-purpose algorithm to analyse the spectra of any planetary atmosphere at any wavelength – from the ultraviolet to the microwave – and under a wide variety of observing geometries – including nadir-viewing, limb-viewing and surface-based observations. Thanks to its versatility, the NEMESIS algorithm has been used to analyse the spectra of most Solar System planets, moons and exoplanets (e.g., [2-8]), made from many different ground and space-based observatories and planetary missions.

The NEMESIS code is written in Fortran and, as a compiled language, provides highly efficient numerical calculations, allowing a fast analysis of the planetary spectra. The high accuracy and versatility in the radiative transfer calculations can translate into some shortcomings in other aspects. In terms of usability, the learning curve can be steep for newcomers, who need to get used to all the flags and capabilities of NEMESIS and a variety of specific formats for the input files, which can complicate the pre-processing and post-processing of the code results. In addition, the complexity of the code presents a significant challenge when implementing new functionality. In terms of performance, while parallel processing can be implemented in Fortran programs, its implementation is often more tedious than in other programming languages. In addition, the static declaration of arrays in NEMESIS can require unused memory in the runs that might prevent the scalability of the code to multiple parallel processes.

Here we present archNEMESIS, a Python implementation of NEMESIS devoted to providing all the capabilities of the original code, while benefiting from the diversity of libraries and packages written in this language to improve usability and portability. The development of archNEMESIS was driven by two main goals, providing the capabilities of NEMESIS in terms of radiative transfer calculations, but also for existing users of NEMESIS to use it as a tool for the pre-processing or post-processing of atmospheric retrievals from the original code.

In this article, we present an overview of the software's architecture, implementation and validation tests. Specific information about how to use the different functionalities of the code is provided in the online software documentation, which can be found at https://archnemesis.readthedocs.io/.

### 1.2. Implementation and architecture

ArchNEMESIS is a code designed to: 1) generate a forward model of the spectrum of any planetary atmosphere based on a set of input parameters; and 2) perform atmospheric retrievals, by which some of the input parameters are iteratively modified until the simulated spectrum best fits the measured one.

The structure of the code can generally be divided into four high-level parts, and the implementation largely relies on the use of Python classes. Python classes are objects that serve as user-defined templates for grouping both data (often called attributes) and the functions (often called methods) that operate on that data. By keeping these components together, classes encourage a clear and modular design that make the code more organised and easier to maintain.

The four high-level parts the code is divided into are:

- Reference classes: In archNEMESIS, all the input information required to model the spectrum of a planet, starting from the atmospheric and surface properties of the planet, and extending to the geometry and specific of the instrument we want to simulate, is stored within several Python classes.

- Model parameterisations: These are a set of parameters that are iteratively modified in our model to find the best fit between the modelled and measured spectrum. In atmospheric retrievals, these variables are more generally known as the state vector.

- Forward model: This refers to the set of functions that solve the radiative transfer equation and allow the calculation of the spectrum based on the model inputs and parameterisations.

- Retrieval engine: This refers to the algorithm or methodology to solve the inverse problem – i.e., search for the set of model parameters that produce a best fit between the modelled and measured spectra.

Figure 1 shows a high-level schematic of the structure of archNEMESIS for a forward model simulation. In the following sub-sections, we describe each of these parts in more detail.

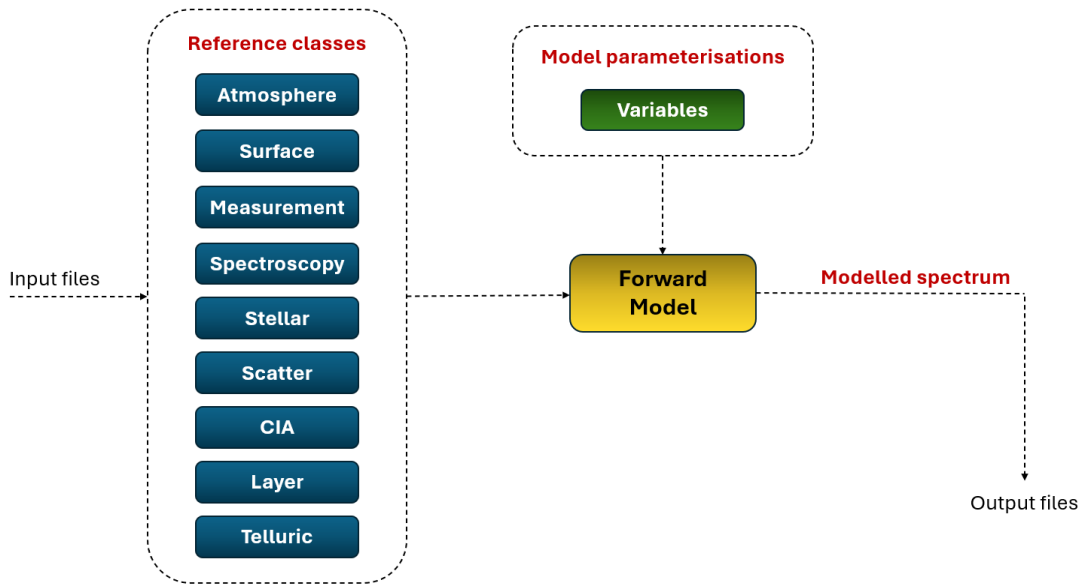

*Figure 1: General structure of archNEMESIS for a forward model simulation. The input information is stored as the attributes of several classes that, together with the model parameterisations we may want to include, carry all the information we need to simulate the spectrum of a planetary atmosphere, which is calculated within the forward model.*

### 1.2.1. Reference classes

In archNEMESIS, all the parameters we need to simulate the electromagnetic spectrum of a planetary atmosphere are stored as the attributes of multiple Python classes. The sub-division of the input information into different classes was performed so that each class represents key aspects of the radiative transfer modelling. This sub-division ensures that classes are independent from each other, allowing users to easily enable or disable them based on their needs. The reference classes hold the information that is fed into the forward model, but they might also be useful for the pre-processing or post-processing of the retrieval. For example, the different classes include methods to easily explore or modify the attributes within the class, perform calculations using these attributes or produce diagnostic plots. At the time of writing of this article, there are nine reference classes to define all the forward model inputs (see Figure 1).

The *Atmosphere* class includes information about the planetary atmosphere we want to model. In particular, it includes information about the planet's basic properties (e.g. mass, radius), and more specifically about its atmosphere, including the reference vertical profiles for the gaseous species (e.g., altitude, pressure, temperature, gas mixing ratios), as well as for the suspended aerosols in the atmosphere (e.g., dust particles, clouds, etc.). For most applications, the atmosphere of a single vertical column of the planet needs to be defined at a specified latitude and longitude. Nevertheless, the *Atmosphere* class also includes the possibility of defining reference vertical profiles at multiple latitudes and longitudes.

The *Surface* class includes the information about the planet's surface, in cases where the planet has a solid surface. In archNEMESIS, the reflection of light from the surface of a planet can be modelled either following a Lambertian reflectance distribution, where the emissivity and albedo are defined, or following the Hapke reflectance model [9], where a set of parameters required for this surface model needs to be defined. Similar to the *Atmosphere* class, the *Surface* class allows the definition of different surface properties at several locations across the planet.

The *Measurement* class includes all the information regarding the instrument setup and measured spectra. In particular, it carries the information about the measured wavelengths and radiances, including the uncertainties in the measurement. In addition, for each measured spectrum, it specifies the relevant observing angles (i.e., emission, azimuth and solar zenith angles), including any information regarding the reconstruction of the field-of-view (FOV). Finally, this class also carries information about the instrument line shape, defining the spectral resolution of the instrument, which can be modelled with any arbitrary function defined by the user.

The *Spectroscopy* class includes information about the absorption cross sections of the gases in the atmosphere. In archNEMESIS, all the radiative transfer calculations are performed using pre-tabulated look-up tables containing the line-by-line cross sections or the *k*-distributions required for solving the radiative transfer equations with the correlated-*k* approximation [10]. In this respect, the *Spectroscopy* class allows the user to easily interact with these tables.

The *Stellar* class includes the information about the reference stellar spectrum, which may or may not be required for specific user needs. For example, it is not needed when modelling only thermal emission calculations from the atmosphere. In addition, it specifies the planet-star distance required to calculate the top-of-atmosphere stellar flux.

The *Scatter* class includes information of two types. Firstly, it includes information about the optical properties of the atmospheric aerosols (i.e., extinction cross section, single scattering albedo, phase function). These optical properties can be calculated using the `Makephase` method of the *Scatter* class following Mie Theory (e.g., [11]). Secondly, in the case of multiple scattering runs, it specifies the parameters for the discretisation of the model into quadrature angles (for the zenith direction) and Fourier components (for the azimuth direction) according to the doubling and adding method [12].

The *CIA* class includes information about the cross sections for modelling the collision-induced absorption (CIA). Similar to the gas absorption cross sections specified in the *Spectroscopy* class, the CIA cross sections are tabulated in look-up tables. In this respect, the *CIA* class allows the user to easily read and write these tables based on the information in the class.

The *Layer* class includes information about the discretisation of the atmospheric vertical profiles into a finite number of layers in which the radiative transfer calculations are performed. The *Layer* class therefore includes information about the number of layers and the methodology used to split these layers (e.g., layers split by equal changes in pressure, or layers split by equal changes in altitude, etc.). This class also includes methods to calculate the average atmospheric properties of the layer.

The *Telluric* class includes information about the Earth's atmosphere in the case that we want to model the spectrum of a planetary atmosphere attenuated by the telluric transmission from the observer's position. To some extent, the *Telluric* class may be thought of as two extra *Atmosphere* and *Spectroscopy* classes defining the properties of the Earth's atmosphere. In addition, this class includes some methods to directly download vertical profiles of the Earth's atmosphere from the ERA5 reanalysis model from the European Centre for Medium-Range Weather Forecasts (ECMWF) at the time and location of the observations [13].

**1.2.2. Model parameterisations**

The reference classes presented in the previous section carry all the information required by the forward model to simulate the spectrum of a planetary atmosphere. However, when performing a retrieval, we generally only want to infer a reduced number of parameters from the inputs and fix all the rest. In addition, the computational time of an atmospheric retrieval quickly increases with the number of fitted parameters. Therefore, the input information in the reference classes is often parameterised to reduce the number of parameters that are iteratively modified in the retrieval – these are known as model parameterisations.

The nature of the model parameterisations is very variable and largely depends on the observed planetary atmosphere and the type of measurement we want to analyse. For example, nadir-viewing observations can have limited sensitivity to the vertical distribution of gaseous abundances and just be sensitive to the integrated column abundance. In that case, a suitable model parameterisation might be a single scaling factor to the gas volume mixing ratio profile listed in the *Atmosphere* class. If using this parameterisation, the retrieval engine will be iterating only one parameter instead of the mixing ratio at each altitude level. Similarly, we may want to use parameterisations to define the vertical structure of clouds – for example, we may parameterise the cloud by defining its bottom level, a fractional scale height and a cloud optical depth. Using such model parameterisations, we can reduce the number of parameters that are iterated in the retrieval.

It must be noted that not necessarily all the model parameterisations apply to the atmosphere, but they may apply to any of the inputs from the reference classes, including other planet properties or even certain characteristics of the instrument. For example, Alday *et al.* [14] used a model parameterisation to retrieve the instrument lineshape from the measured spectra, modelled as a double Gaussian function. In addition, as it will be further detailed in the next section, some model parameterisations may not even modify the input information to the forward model, but might modify the calculated spectra. This option is included in the code to account, for example, for instrumental effects that cannot be easily removed in the calibration of the spectra (e.g., fringe patterns).

In archNEMESIS, the information regarding the model parameterisations is stored in the *Variables* class. Each of the model parameterisations has a unique identifier, which allows their correct implementation. Most of the model parameterisations in archNEMESIS are a heritage of the original NEMESIS code and the numerical identifiers are respected to allow backward compatibility between the two. This part of the algorithm is highly dynamic and new parameterisations are constantly being included by the users of the code. Therefore, we recommend the users of archNEMESIS to check the online documentation to see the latest additions, as well as to see how users can implement their own parameterisations.

**1.2.3. Forward Model**

The forward model refers to the set of functions by which the electromagnetic spectrum of a planetary atmosphere is calculated based on the information from the reference classes and the selected model parameterisations. In archNEMESIS, all these functions are implemented within the *ForwardModel* class. In essence, the attributes of the *ForwardModel* class are the reference classes and the *Variables* class (see Figure 1), and the methods within this class are used to perform the radiative transfer calculations and calculate the spectra.

Figure 2 shows a sketch of the most relevant methods within the standard forward model implementation in archNEMESIS. As mentioned, the attributes of the *ForwardModel* class are the reference classes and the model parameterisations, which are fed into the `subprofretg` method. In this method, the reference classes are modified based on the selected model parameterisations, and the updated classes are returned. After this, the `calc_path` method of the *ForwardModel* class splits the atmosphere into a finite number of layers and calculates the atmospheric paths based on the geometry of the observation - i.e., for each layer, it determines the mean temperature and pressure, as well as the integrated density along the direction of the line of sight. `CIRSrad` is the main method where the radiative transfer calculations are performed, first calculating the relevant optical properties of each layer (e.g., absorption and scattering optical depths), solving the multiple scattering field if required [11], and finally calculating the electromagnetic spectrum. After the spectrum has been calculated, the `subspecret` method modifies the spectrum based on any relevant model parameterisations that might be added to mimic some instrument characteristics. Finally, the spectrum is convolved with the instrument lineshape to account for the spectral resolution of the instrument.

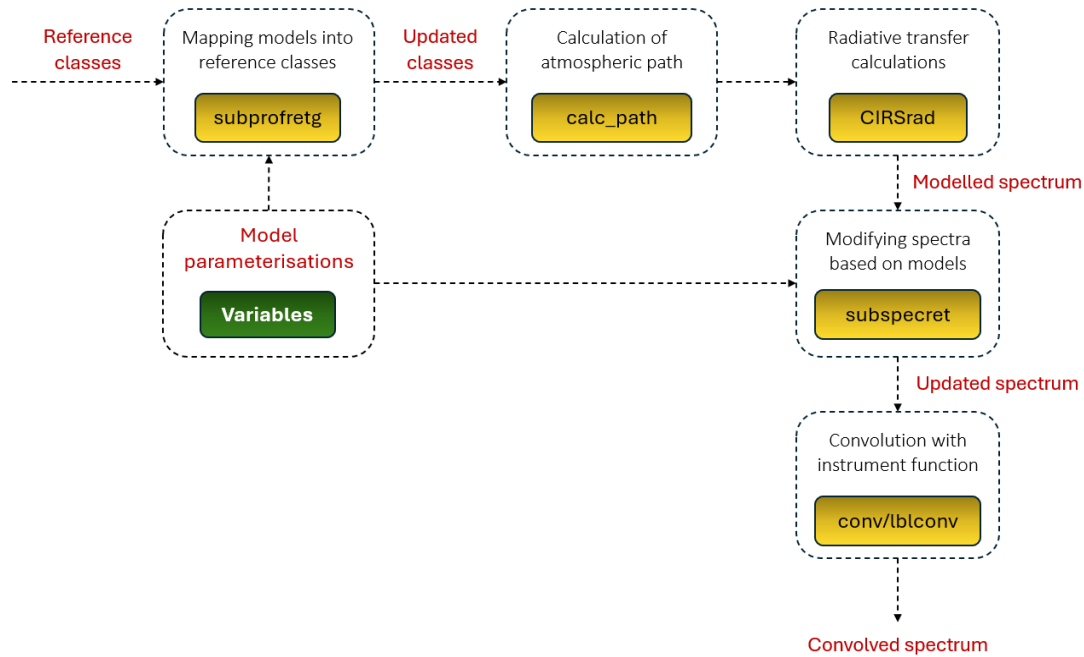

*Figure 2: General structure of the forward model. The attributes of the ForwardModel class, given by the reference classes, are first updated based on the model parameterisations. Using these updated attributes, the code calculates the atmospheric paths and performs the radiative transfer calculations, simulating the spectrum of the planetary atmosphere we want to model. The modelled spectrum is then updated by any relevant model parameterisations and is finally convolved with the instrument function to simulate the specifics of the instrument we are modelling.*

The implementation presented in Figure 2 describes the standard forward model in archNEMESIS, where a single spectrum of the planet needs to be computed. For some applications, several spectra of the same atmospheric column may need to be computed and, in this case, some optimisations can be performed to increase the speed of the code. In archNEMESIS, these special implementations of calculations are generally introduced in independent methods of the *ForwardModel* class, but always following the same logic as presented in Figure 2.

For example, *nemesisSO* is the name of the forward model optimised for solar occultation observations, where spectra of the same atmospheric column need to be computed at a set of tangent heights (since spherical symmetry is assumed). In this case, all the atmospheric paths are simultaneously computed, so that most of the calculations are only performed once (e.g., calculation of the optical properties of each atmospheric layer), improving the computational efficiency by a substantial factor.

Similarly, the *nemesisC* version of the forward model increases the speed of the calculation in a similar fashion, but valid for nadir-viewing or upward-looking geometries with full multiple scattering calculations. In this version, we may specify several simultaneous different geometries for observing the same atmospheric column, a valid assumption when observing the same location of a planet by several spacecraft or, for example, when a single instrument observes a planet's disk with constant atmospheric properties across certain regions (e.g., longitudinally-symmetric). In this version, the multiple scattering calculations, which are the most computationally-expensive part of the computations, are calculated only once, drastically increasing the efficiency of the code.

In the future, different needs or observing modes may benefit from the implementation of specific modifications of the forward model as in the examples above. Thus, we recommend that users of archNEMESIS check the documentation to see the latest additions or reach out for the implementation of specific needs.

### 1.2.4. Retrievals

Apart from performing the radiative transfer calculations to compute the electromagnetic spectrum of a planetary atmosphere, archNEMESIS includes a retrieval engine to solve the inverse problem, by which we try to find the atmospheric properties that produce a best match between the modelled and measured spectrum.

This retrieval engine sits on top of the rest of the parts of the code explained in the previous sections. In particular, similar to the forward model, the inputs to the retrieval engine are the reference classes and the selected model parameterisations (see Figure 3). The retrieval engine then computes a spectrum using the forward model and, based on the differences between the measured and modelled spectra, decides how the model parameters should be adjusted to minimise the difference between the two. After several iterations the retrieval converges to within preset conditions and the retrieval engine returns the model parameters that produce a best fit between the modelled and measured spectra, the estimated uncertainties in the parameters, and the best-fit modelled spectrum.

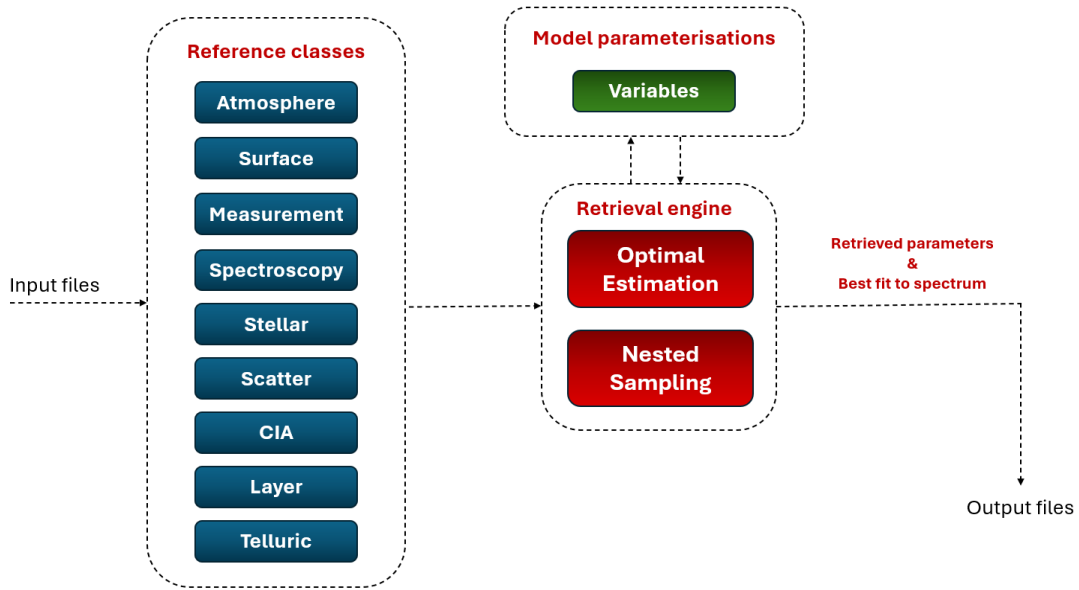

*Figure 3: General structure of the archNEMESIS for a retrieval run. Similar to the forward model scenario, the reference classes and model parameterisations provide all the necessary information to run the simulation. The retrieval engine runs the forward models and modifies the model parameters to find the set that produce a best fit between the measured and modelled spectra. The retrieval engine will provide as an output the best fit to the measurement, as well as the values of the model parameters that produce the best fit.*

Inverse problem solvers have been studied for decades and several different approaches might be used, depending on specific needs, such as non-linear least squares methods or Markov chain Monte Carlo (MCMC) algorithms. ArchNEMESIS allows the execution of the retrieval using two widely used algorithms in planetary science – the optimal estimation and the nested sampling methods.

Optimal estimation is a widely used Bayesian-based technique to solve the inverse problem in atmospheric retrievals [15]. In particular, the optimal estimation algorithm aims to find a set of model parameters that minimise the difference between the measured and modelled spectra, subject to a minimal departure from the prior information about the atmosphere. It is an iterative approach in which the retrieval gradually converges until it finds the minimum in a cost function. The specific algorithm implemented in archNEMESIS is equivalent to that described by Irwin *et al.* [1] for NEMESIS.

On the other hand, nested sampling is a Bayesian inference technique that is widely used in the analysis of spectra from exoplanetary atmospheres. It is a Monte Carlo-based method that performs a much more exhaustive map of the parameter space than iterative approaches such as optimal estimation, allowing it to handle multi-modal posterior distributions of the fitted parameters, as well as strong degeneracies between these. The implementation of this retrieval engine within archNEMESIS is performed through the usage of the *pymultinest* library [16], which itself relies on the *MultiNest* nested sampling algorithm [17].

Both retrieval techniques are highly useful for the analysis of planetary atmospheric spectra, both having their advantages and disadvantages. In terms of computational speed, the optimal estimation method is much faster, typically only needing a few iterations to reach a solution, each requiring the computation of N+1 forward models. In contrast, since nested sampling explores the parameter space in more depth, it requires a large number of forward models to reach a solution, and the number of forward models can potentially scale exponentially with the number of fitted parameters, making it less suitable for high-dimensional problems. In terms of convergence, the more exhaustive exploration of the parameter space makes nested sampling a more robust algorithm, being able to reproduce complex distributions and degeneracies. The convergence of the optimal estimation algorithm might be difficult in such complex cases, where the method might get stuck in a local minimum of the cost function.

Running both types of retrievals with archNEMESIS is relatively straightforward, since both essentially require similar input information. In particular, the initial set of parameters in the *Variables* class, as well as the prior uncertainties, are defined in the .apr file (see section 1.2.5), and the retrieval engine can just be selected by setting a flag. We recommend that users check the documentation to see specific examples on how to run a retrieval using each method.

### 1.2.5. Input files

The information required by archNEMESIS to run either a forward model or a retrieval needs to be specified in a set of input files with a specific format. In particular, there are two versions of the files that can be used, one using the existing NEMESIS input files, and another one specifically designed for archNEMESIS.

ArchNEMESIS includes several functions and methods to read and write the input and output files with the NEMESIS format. In particular, NEMESIS reads the input information from text files with the same name but different extensions (e.g., .ref, .sur, .set, etc.) and, after the retrieval or forward model has terminated, writes out some other text files with other extensions (e.g., .mre, .cov). In archNEMESIS, there are several functions to read and write these files and fill the attributes within the reference classes. This option was implemented to ensure backward compatibility between the two codes. In addition, this option may also prove useful for users that prefer to run the retrievals with NEMESIS but want to use the archNEMESIS classes for the pre-processing and post-processing of the retrievals.

In addition to the standard NEMESIS files, we have created a version of the input files specific for archNEMESIS. In particular, a forward model or retrieval run in archNEMESIS can be performed by reading information from a single HDF5 file. This file format allows the organisation of the data in a hierarchical manner using groups and datasets. This format is particularly useful for the structure of the inputs from archNEMESIS, since each of the reference classes corresponds to a unique group within the file, which includes all the information (i.e., datasets) for filling the attributes of that particular class. Therefore, the interaction between the input file and each of the classes is very straightforward – all classes have `read_hdf5` and `write_hdf5` methods to easily read/write the information. In addition, the HDF5 files allow us to include metadata explaining the meaning of each of the attributes within the class, making the input files more easily readable.

In both versions of the input files, the information regarding the model parameterisations is read from a text file with a .apr extension. This file format is heritage from the standard NEMESIS code and in this case is also preferred over an HDF5 file because of the high variability of the file format depending on the selected model parameterisations; each model parameterisation requires different information for its correct implementation within the code, which in turn requires the information in the input .apr file to be specified in a different format. The format required for each model parameterisation implemented in archNEMESIS is indicated in the online documentation of the code, which will be updated as new parameterisations are included.

**1.2.6. Code optimisation**

Apart from improving the usability and accessibility of the code, one of the main goals of archNEMESIS is to provide all the capabilities of NEMESIS at a similar or exceeding computational speed. Given that the original NEMESIS code is written in Fortran, which is by nature more efficient than Python at performing numeric calculations, we have included certain optimisation techniques to increase the efficiency of archNEMESIS.

One of the reasons why Fortran is substantially faster than Python is because the latter is an interpreted language, meaning that each line of the code is processed by the Python virtual machine at runtime, which makes it intrinsically slower than Fortran, which is a compiled language. In order to increase the speed of archNEMESIS, the most computationally-expensive parts of the code (e.g., multiple scattering calculations or *k*-distribution combinations) are compiled using *numba*'s just-in-time (JIT) compilers [18]. These compilers translate Python into machine code, removing a substantial part of the overhead Python introduces, and providing computational speeds comparable to C/Fortran. One point to note is that *numba* compiles functions at runtime the first time they are called, meaning there can be a slight delay on the initial execution. However, once compiled, the functions will run at near-native speeds.

Additionally, the archNEMESIS reference classes are highly versatile, allowing users to enable or disable them as needed. This flexibility ensures that only the strictly necessary calculations are performed, enabling efficient modelling of the electromagnetic spectrum from any planetary atmosphere.

Finally, the speed of archNEMESIS retrievals is substantially increased by the use of parallel processing. In particular:

- Optimal estimation retrievals require the computation of the Jacobian matrix in each iteration, which defines the partial derivatives of the spectrum with respect to each of the fitted parameters in the model. For non-scattering calculations these partial derivatives can be calculated analytically for most atmospheric parameterisations, making the calculations highly efficient. However, the mathematical complexity of multiple scattering calculations requires the Jacobian matrix to be computed numerically by successively perturbing the values of the fitted parameters and recording the differences in the forward-modelled spectra. This means that for each iteration, N+1 forward models need to be computed, N being the number of fitted parameters. Since the calculations are independent, these may be run in parallel by archNEMESIS for multiple-scattering calculations, drastically increasing the speed of the code by a factor dependent on the number of parallel processes called.

- Nested sampling retrievals require the computation of a large number of forward models, often several orders of magnitude larger than what is required for optimal estimation retrievals. Similar to archNEMESIS's calculation of the Jacobian matrix for multiple-scattering optimal estimation retrievals, these independent evaluations of the forward model are easily parallelised, increasing the speed of the code by a factor dependent on the number of parallel processes called.

## 1.3. Quality control

### 1.3.1. Automated tests

As part of the development of archNEMESIS, a suite of automated tests has been implemented using the *pytest* framework. These tests are designed to ensure that new features and updates do not break existing functionality, while also confirming that the code behaves correctly under a variety of scenarios. ArchNEMESIS is connected to GitHub Actions to automatically run the *pytest* suite on every commit or pull request, ensuring that no existing functionalities are broken in the repository.

### 1.3.2. Input checks

The versatility in the calculations provided by archNEMESIS requires the correct setup of flags and variables, which are mostly specified in each of the reference classes and input files. In order to avoid the presence of conflicting flags, and ensure that all required information is present based on the specified flags, all reference classes include a method to check the validity of the input setup. These methods are automatically applied when reading/writing from/to the input files to ensure the calculations are logical within the code's framework.

**1.3.3. Validation tests**

Ample exercises have been performed to validate the radiative transfer calculations of archNEMESIS with respect to the original NEMESIS code, both in terms of forward modelling and retrieval convergence. In the online software documentation, there are several jupyter notebooks including examples of the different capabilities of archNEMESIS, which include in some cases direct comparisons with the outputs of the original NEMESIS version [1] in a variety of forward model scenarios (e.g., no scattering and multiple scattering scenarios). Similarly, these jupyter notebooks include some validation tests against the results from the DISORT radiative transfer code [19], which provides further evidence for the accuracy and validity of the code.

**(2) Availability**

*2.1. Operating system*

Works in all operating systems supporting Python.

*2.2. Programming language*

Python 3.9 or higher.

*2.3. Dependencies*

*ArchNEMESIS* is fully written in Python and built upon other several open-source libraries. In particular:
- Numerical calculations: *numpy; scipy.*
- Plotting: *matplotlib, basemap.*
- File handling: *h5py*.
- Optimisation: *numba; joblib*.
- Nested sampling: *pymultinest*.
- Extraction of ERA-5 model profiles: *cdsapi; pygrib*.
- Testing: *pytest*.

*2.4. List of contributors*

Here, we list the key contributors to archNEMESIS (AN) and NEMESIS (N), code from which the archNEMESIS is built.

**Juan Alday (AN):** The Open University, Milton Keynes, UK.
**Joseph Penn (AN):** University of Oxford, Oxford, UK.
**Patrick Irwin (N):** University of Oxford, Oxford, UK.
**Jonathon Mason (AN):** The Open University, Milton Keynes, UK.
**Jingxuan Yang (AN):** University of Oxford, Oxford, UK.
**Leigh Fletcher (N):** University of Leicester, Leicester, UK.
**Nick Teanby (N):** University of Bristol, Bristol, UK.
**Jo Barstow (N):** The Open University, Milton Keynes, UK.
**Conor Nixon (N):** NASA Goddard Spaceflight Center, Greenbelt, MD, USA.

## 2.5. Software location:

### Archive
**Name:** Zenodo
**Persistent identifier:** https://doi.org/10.5281/zenodo.14746548
**Licence:** GNU General Public License v3.0
**Publisher:** Juan Alday
**Version published:** 1.0.0
**Date published:** 27/01/2025

### Code repository
**Name:** GitHub
**Identifier:** https://github.com/juanaldayparejo/archnemesis-dist
**Licence:** GNU General Public License v3.0
**Date published:** 26/06/2024

## 2.6. Language

All documentation is provided in English.

## (3) Reuse potential

ArchNEMESIS can be used by researchers both within planetary science and from related fields in several ways. For instance, its retrieval capabilities can be used to analyse the spectra of Solar System planets and exoplanets measured by satellites or from ground- and space-based telescopes. Similarly, its forward modelling capabilities can provide aid in the design of instrumentation for future space missions. Additionally, archNEMESIS can be used as a pre- and post-processing tool for the widely used NEMESIS radiative transfer algorithm, thereby being useful to those already relying on NEMESIS. Because of its modular design, users can adapt or extend it to accommodate new model parameterisations and specific forward-model setups for different scenarios. Contributors are encouraged to discuss their ideas by reaching out via the project's issue tracker or through direct email, ensuring that any updates or modifications benefit the broader user community.

For researchers focused on exoplanet studies, we also suggest exploring NEMESISPY [20], another implementation of the NEMESIS code specialised for exoplanet work.

In terms of ongoing support, we intend to actively maintain the software so that it continues to incorporate state-of-the-art radiative transfer techniques and remains compatible with emerging datasets and instrumentation requirements. While we aim to address queries in a timely manner, users should be aware that the level of direct support may vary depending on available resources. Nevertheless, the archNEMESIS' open development philosophy (with publicly accessible source code and detailed documentation) provides a robust foundation for collaborative feature development.

**Acknowledgements**
The authors thank the many developers of the NEMESIS radiative transfer and retrieval tool, as well as the developers of open-source Python libraries that archNEMESIS is built upon.

**Funding statement**
We acknowledge support from the Science and Technology Facilities Council (ST/Y000234/1, ST/X001180/1).

**Competing interests**
The authors declare that they have no competing interests.